# Boundary effect on acoustic cloak with unideal pentamode material

Yi Chen, Xiaoning Liu[*] and Gengkai Hu

Key Laboratory of Dynamics and Control of Flight Vehicle, Ministry of Education, School of Aerospace Engineering, Beijing Institute of Technology, Beijing 100081, China

Correspondence: liuxn@bit.edu.cn

**Abstract**

Pentamode materials are elastic solids with vanishing shear modulus, and can be used cloak underwater sound with solid state and broadband merits. However, pentamode materials realized with real microstructure have inevitable small shear modulus. This paper systematically studies the impact of shear rigidity and inner surface constraints on acoustic cloak with unideal pentamode material. The shear rigidity introduces a new kind of resonance in the radial direction, which is different from the traditional whispering-gallery resonance along the circumference. Totally fixed, radially fixed or free boundaries on the cloak inner surface are found to show significant difference on the cloaking function. To realize a broadband cloak with suitable boundary for practice, we propose to attach an elastic thin shell on the inner surface to virtually tune the boundary constraint. The proposed strategy is also validated with microstructure cloak simulation. This study will provide valuable guidance for the practical applications of pentamode acoustic cloak.

**Key Words**: pentamode acoustic cloak; resonance; damping; boundary condition; microstructure cloak; broadband

## 1. Introduction

Invisible cloak is an ultimate example of controlling wave through material distributions, for which the transformation approach provides a general tool to find the required material parameter [1,2]. As for acoustics cloaks, meta-fluids with anisotropic densities were initially proposed in the same line with the transformation optics [3,4], namely the inertial cloak [5]. A variety of meta-fluids have been proposed for realization of anisotropic densities such as alternating fluid layers [6], perforated plates [7] or resonant schemes [8,9]. However, these techniques require fluids as working media or suffer from narrow frequency band, thus their engineering implications are limited. Further, the achievable density anisotropy is quite small. The two principal densities can differ by five times for air sound [10], while only differ by two times for water sound [11]. Therefore, only the carpet cloak [12,13] was experimentally demonstrated so far due to much larger anisotropy is required for omnidirectional cloaks.

Pentamode materials (PM), proposed by Milton and Cherkaev in 1995 [14], are degenerated elastic solids with rather high achievable anisotropy in modulus. These materials cannot resist shear deformation and acoustically resemble ordinary fluids with their moduli however being anisotropic. In 2008, Norris proved that PM is also capable of controlling acoustic waves through transformation method [5]. By careful microstructural design to tune the shear rigidity, PM can be approximated by solid materials. The greatest advantage of PM is that, unlike many other metamaterials, they are not resonance based and are intrinsically broadband. Other advantages include sharper control benefitting from higher anisotropy, and the solid nature of controlling devices. These merits stimulated intense researches on PM recently, e.g., the PM transformation theory [15,16], acoustic wave controlling applications [17-27], and PM microstructure design [28-34]. A PM acoustic cloak in cylindrical configuration with concrete microstructures was designed and numerically verified by Chen et al. [19], and was further experimentally tested to demonstrated its manipulation capability for underwater sound [26].

There are several facts preventing a real cloak from being perfectly invisible over broadband frequency. Reasons can be that the cloak is not mapped out from an infinitesimal point in the virtual space, that the smoothly gradient cloak is discretized by piecewise layers, or that the homogenization of the metamaterial is not accurate. For PM cloaks, an additional harmful fact is that the practical solid-based PM cannot be ideal, i.e. with zero shear resistance, in order to be statically stable. A real PM in fact falls into the category of orthotropic solids. For example, the ratio of shear to bulk modulus are usually about 1% for 2D PM [19], or are about 1‰ for 3D PM [35], but not zero. The imperfectness of PM is a special issue compared to other facts, because it induces extra shear modes in contrast to longitudinal mode in ideal PM. It has been numerically found that, a real PM cloak shows shear resonance scattering instead of being broadband as expected for ideal PM [19]. Smith and Verrier [36] also studied an acoustic cloak composed of solids with anisotropic density but isotropic modulus, and found similar shear resonance. However, their study is apparent different from the current one based on PM with anisotropic modulus. Despite the resonances can be suppressed by material damping, the cloak inner surface boundary is another important factor for cloaking effect. In many previous studies [15,23,24], a radially fixed inner surface boundary is in general assumed to demonstrate the best invisibility. However, it should be emphasized here that, the radially fixed boundary is only reasonable for air sound, since conventional solids, e.g., plastic or metal, can be used as acoustic rigid material. As for elastic wave in the PM cloak, ordinary solids easily couple with the PM and can hardly restrict the motion of the inner surface. From a practical point of view, the most realizable boundary for implementation is a free inner surface. As will be shown, however, the free inner surface significantly influences the cloaking performance, and a practical solution is demanding to obtain a promising invisibility. Due to much complex wave propagation in gradient unideal PM, most analytical contributions consider the PM as ideal ones with zero shear rigidity [22,23], and very limited attention has been paid to the impact of imperfectness of PM as well as the boundary effect on the acoustic cloak.

In this work, we first derive a semi-analytical model of the acoustic cloak with unideal PM, and then systematically investigate the impact of material parameters and inner surface boundary on the cloaking performance. The paper is organized as follows. In Section 2, the PM and cylindrical cloak is described and two parameters are introduced to quantify the PM imperfectness. Section 3 is concerned with the derivation of the scattering solution for unideal cylindrical PM acoustic cloak. In Section 4, the impact of material parameters and boundary conditions on the cloaking performance are investigated thoroughly. A practically feasible boundary is further proposed to achieve a broadband cloaking performance and is verified through microstructure cloak. Finally concluding remarks are given in Section 5.

## 2. Characterization of pentamode material and model of cylindrical cloak

Pentamode materials are defined as elastic material with five zero eigenvalues of its elasticity matrix, the eigenvector **S** corresponding to the single non-zero eigenvalue is the only stress the material can bear. The elasticity tensor can be expressed as $\mathbf{C} = K\mathbf{S} \otimes \mathbf{S}$, and the stress is proportional to the characteristic tensor **S**, $\boldsymbol{\sigma} = p\mathbf{S}$, with $p$ being termed as pseudo pressure. The trivial isotropic case with $\mathbf{S} = \mathbf{I}$ represents ordinary fluids without shear rigidity. In principal coordinate system, the elasticity matrix of a designed 2D PM has inevitable shear modulus and takes the following form [19]

$$\mathbf{C} = \begin{pmatrix} K_x & K_{xy} & 0 \\ K_{xy} & K_y & 0 \\ 0 & 0 & G_{xy} \end{pmatrix}. \tag{1}$$

Two parameters characterizing the degree of imperfectness can be defined as

$$v = \frac{|K_{xy}|}{\sqrt{K_x K_y}}, \qquad \mu = \frac{G_{xy}}{\sqrt{K_x K_y}}. \tag{2}$$

In order to approximate a perfect PM, both $|v - 1| \ll 1$ and $\mu \ll 1$ are required. For isotropic material case, requiring $v \ll 1$ is enough since $v = 1 - 2\mu$. For anisotropic materials, both $v$ and $\mu$ should be restricted.

The material parameter for PM cloak can be derived from transformation theory as shown in Fig. 1 [5]. First, we suppose a virtual space $\Omega$ & $\Omega^{out}$ occupied by homogenous fluid with density $\rho_0$ and modulus $K_0$. Second, through a coordinate mapping $\mathbf{x} = f^{-1}(\mathbf{X})$, one can map the acoustic equation in $\Omega$ onto PM acoustic equation in $\omega$. Acoustic wave along a straight trajectory in the virtual space (Fig. 1) will travel along a curved route in the cloak and leave the central region undetectable. The required density and elastic matrix of PM in the cloak region $\omega$ is then derived as [5]

$$\rho' = \rho'\mathbf{I}, \quad \mathbf{C} = \begin{pmatrix} K_r & \sqrt{K_r K_\theta} & 0 \\ \sqrt{K_r K_\theta} & K_\theta & 0 \\ 0 & 0 & 0 \end{pmatrix} \quad (3)$$

$$\rho' = \rho_0 \frac{f(r)f'(r)}{r}, \quad K_r = K_0 \frac{f(r)}{rf'(r)}, \quad K_\theta = K_0 \frac{rf'(r)}{f(r)}, \quad (4)$$

where $K_r$ and $K_\theta$ are the two principal moduli along the radial and circumferential directions, respectively. Here, different functions $f(r)$ can be used to simplify the material parameters provided that $f(b) = b$ and $f(a) = \delta$, where $a$ and $b$ are the inner and outer radius of the cloak shell (Fig. 1(b)). A small radius $\delta \ll 1$ is adopted to avoid the material singularity. Among the choices of $f(r)$, the commonly used one leading to uniform density or modulus can be unified in a power form

$$f(r) = \left( \frac{b^n - \delta^n}{b^n - a^n} r^n - \frac{a^n - \delta^n}{b^n - a^n} b^n \right)^{1/n}. \quad (5)$$

The linear mapping, uniform density mapping and uniform modulus mapping can be obtained by taking $n$ to be 1, 2 and $+\infty$, respectively.

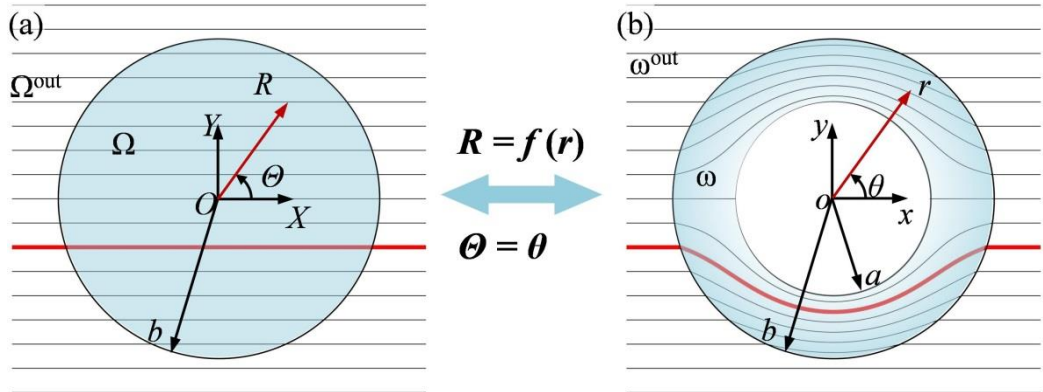

**Fig. 1.** Demonstration of transformation approach for pentamode cloak; (a) Virtual space occupied with homogenous fluid; (b) Physical space with background region $\omega^{out}$, cloak region $\omega$; $\Omega$ and $\omega$ have the same outer boundary.

3. **Scattering of the cloak with imperfect pentamode material**

In this section, we will address the scattering of a cylindrical cloak with imperfect PM, i.e., essentially a cloak shell of orthotropic solids with gradient density $\rho(r)$ and modulus $\mathbf{C}(r)$ with principal axis aligned with polar frame. A closed form solution for the acoustic scattering of a cylindrical orthotropic solid shell will be very tedious if not impossible. Here, the semi-analytical state space approach traditionally used for laminated orthotropic plates [37,38] is adopted. The original cloak is first discretized into $N$ layers (Fig. 2(b)), with each layer being sufficiently thin for reasons explained below. For the $j^{th}$ layer enclosed by radius $r_{j-1}$ and $r_j$, the PM material is treated as homogenous with density and modulus evaluated at the middle point, $\rho_j = \rho_j((r_{j-1} + r_j)/2)$, $\mathbf{C}_j = \mathbf{C}_j((r_{j-1} + r_j)/2)$. The background domain outside the cloak is filled with a homogeneous fluid ($\rho_0$, $K_0$), and is denoted as the $(N + 1)^{th}$ layer.

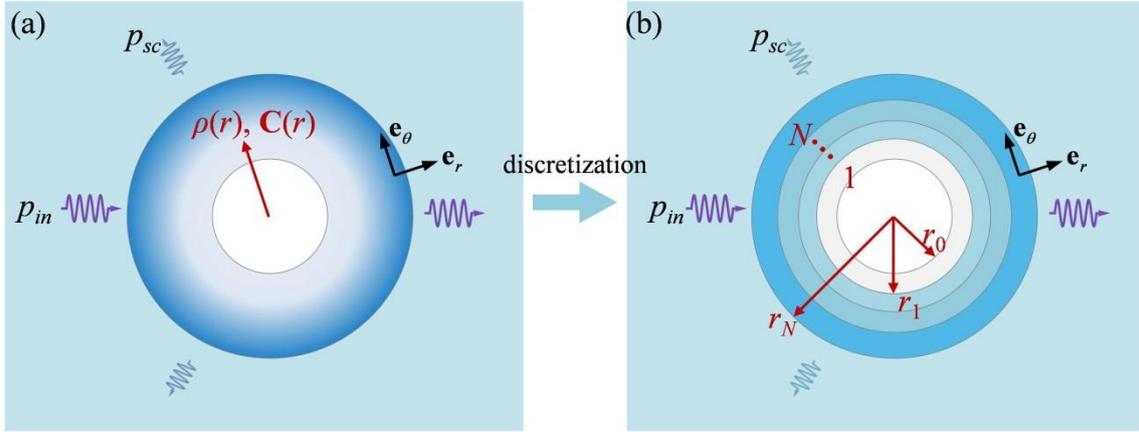

**Fig. 2.** Scattering model of pentamode cloak; (a) A cloak with continuous material parameter is discretized into (b) a cloak with $N$ layers of orthotropic material.

Consider a plane wave with circular frequency $\omega$ incidents from the left, $p_{in} = \exp(ik_0 x)$ with the time dependence term $\exp(-i\omega t)$ omitted for simplification. The incident plane wave can be decomposed as following

$$p_{in} = \sum_{n=0}^{\infty} a_n J_n(k_0 r) \cos n\theta, \qquad (6)$$

where $k_0 = \omega/c_0$ is the wave number in the background, $c_0 = (K_0/\rho_0)^{1/2}$ is the wave speed, $J_n(k_0 r)$ is the $n^{th}$ order Bessel function. The incident coefficient is $a_n = (2 - \delta_{0n})i^n$ ($n > 0$) with $\delta$ representing the Kronecker delta. Scattered pressure $p_{sc}$ in the background fluid is governed by the wave equation, $\nabla^2 p_{sc} + (k_0)^2 p_{sc} = 0$, and follows the following decompositions

$$p_{sc} = \sum_{n=0}^{\infty} b_n H_n^{(1)}(k_0 r) \cos n\theta, \qquad (7)$$

where $b_n$ is the unknown $n^{\text{th}}$ order scattering coefficient, $H_n^{(1)}(k_0 r)$ is the $n^{\text{th}}$ order Hankel function of the first kind. To solve the scattering coefficient, the wave propagation in each PM layer should be considered. For the $j^{\text{th}}$ layer, the constitutive, geometry and momentum equations in the polar coordinate can be expressed as

$$\begin{pmatrix} \sigma_{jr} \\ \sigma_{j\theta} \\ \sigma_{jr\theta} \end{pmatrix} = \begin{pmatrix} K_{jr} & K_{jr\theta} & 0 \\ K_{jr\theta} & K_{j\theta} & 0 \\ 0 & 0 & G_{jr\theta} \end{pmatrix} \begin{pmatrix} \varepsilon_{jr} \\ \varepsilon_{j\theta} \\ 2\varepsilon_{jr\theta} \end{pmatrix}, \tag{8}$$

$$\varepsilon_{jr} = \frac{\partial u_{jr}}{\partial r}, \quad \varepsilon_{j\theta} = \frac{1}{r}\frac{\partial u_{j\theta}}{\partial \theta} + \frac{u_{jr}}{r}, \quad \varepsilon_{jr\theta} = \frac{1}{2}(\frac{\partial u_{j\theta}}{\partial r} + \frac{1}{r}\frac{\partial u_{jr}}{\partial \theta} - \frac{u_{j\theta}}{r}), \tag{9}$$

$$\frac{\partial \sigma_{jr}}{\partial r} + \frac{1}{r}\frac{\partial \sigma_{jr\theta}}{\partial \theta} + \frac{\sigma_{jr} - \sigma_{j\theta}}{r} = -\rho_j \omega^2 u_{jr}, \quad \frac{\partial \sigma_{jr\theta}}{\partial r} + \frac{1}{r}\frac{\partial \sigma_{j\theta}}{\partial \theta} + \frac{2\sigma_{jr\theta}}{r} = -\rho_j \omega^2 u_{j\theta}, \tag{10}$$

where $K_{jr}$, $K_{j\theta}$, $G_{jr\theta}$ are the moduli of the $j^{\text{th}}$ layer, $\varepsilon_{jr}$, $\varepsilon_{j\theta}$, $\varepsilon_{jr\theta}$ are the strains, and $\sigma_{jr}$, $\sigma_{j\theta}$, $\sigma_{jr\theta}$ are the corresponding stresses. Considering the orthogonality of the decomposed mode of different orders and the continuity requirement at the PM/fluid interface, the following ansatz for displacements and stresses are adopted

$$u_{jr} = \sum_{n=0}^{\infty} u_{jnr}(r)\cos n\theta, \quad u_{j\theta} = \sum_{n=0}^{\infty} u_{jn\theta}(r)\sin n\theta, \tag{11}$$

$$\sigma_{jr} = \sum_{n=0}^{\infty} \sigma_{jnr}(r)\cos n\theta, \quad \sigma_{jr\theta} = \sum_{n=0}^{\infty} \sigma_{jnr\theta}(r)\sin n\theta, \quad \sigma_{j\theta} = \sum_{n=0}^{\infty} \sigma_{jn\theta}(r)\cos n\theta, \tag{12}$$

where $(u_{jnr}, u_{jn\theta})$ and $(\sigma_{jnr}, \sigma_{jnr\theta}, \sigma_{jn\theta})$ are denoted as the $n^{\text{th}}$ order displacement and stress components in the polar coordinates system, respectively, in the $j^{\text{th}}$ layer. To be responsible for the interface continuous condition, the $n^{\text{th}}$ order state space vector is defined as,

$$\mathbf{D}_{jn}(r) = \{u_{jnr}(r), u_{jn\theta}(r), \sigma_{jnr}(r), \sigma_{jnr\theta}(r)\}^{\text{T}}. \tag{13}$$

Substituting the above mode expansions Eq. (11) and Eq. (12) into Eq. (10), and after some algebra one can derive a coupled ordinary differential equation for the $n^{\text{th}}$ order state vector of the $j^{\text{th}}$ layer in matrix form

$$\frac{\text{d}}{\text{d}r}\mathbf{D}_{jn}(r) = \mathbf{P}_{jn}(r)\mathbf{D}_{jn}(r), \tag{14}$$

where the matrix $\mathbf{P}_{jn}(r)$ takes the following form

$$\mathbf{P}_{jn}(r) = \begin{pmatrix} -\dfrac{1}{r}\dfrac{K_{jr\theta}}{K_{jr}} & -\dfrac{n}{r}\dfrac{K_{jr\theta}}{K_{jr}} & \dfrac{1}{K_{jr}} & 0 \\ \dfrac{n}{r} & \dfrac{1}{r} & 0 & \dfrac{1}{G_{jr\theta}} \\ \dfrac{1}{r^2}\xi_j - \rho_j\omega^2 & \dfrac{n}{r^2}\xi_j & \dfrac{1}{r}(\dfrac{K_{jr\theta}}{K_{jr}}-1) & -\dfrac{n}{r} \\ \dfrac{n}{r^2}\xi_j & \dfrac{n}{r^2}\xi_j - \rho_j\omega^2 & \dfrac{n}{r}\dfrac{K_{jr\theta}}{K_{jr}} & -\dfrac{2}{r} \end{pmatrix} \quad (15)$$

and $\xi_j = (K_{jr}K_{j\theta} - (K_{jr\theta})^2)/K_{jr}$. The above treatment transforms the original second order partial differential equation into a first order ordinary differential equation. Though the material in the $j^{\text{th}}$ PM layer is homogeneous, Eq. (14) cannot be solved explicitly because $\mathbf{P}_{jn}(r)$ depends on location $r$. To this end, we discretize the cloak into sufficiently thin layers so that $\mathbf{P}_{jn}(r)$ can be considered as a constant matrix evaluated at the midpoint of the layer. As such, Eq. (14) admits an exponential solution, and the two state space vectors at the front and back surface of the $j^{\text{th}}$ PM layer are related by,

$$\mathbf{D}_{jn}(r_j) = \exp\left((r_j - r_{j-1})\mathbf{P}_{jn}(\dfrac{r_{j-1}+r_j}{2})\right)\mathbf{D}_{jn}(r_{j-1}). \quad (16)$$

Using the continuity of the state space vectors at the interface of adjacent PM layers, i.e., $\mathbf{D}_{jn}(r_{j-1}) = \mathbf{D}_{(j-1)n}(r_{j-1})$, the state space vector at the outer surface of cloak, $\mathbf{D}_{Nn}(r_N)$, relates to that of the inner most surface, $\mathbf{D}_{1n}(r_0)$, by a transmittance relation

$$\mathbf{D}_{Nn}(r_N) = \mathbf{T}_n \mathbf{D}_{1n}(r_0), \quad \mathbf{T}_n = \prod_{j=1}^{j=N} \exp\left((r_j - r_{j-1})\mathbf{P}_{jn}(\dfrac{r_{j-1}+r_j}{2})\right). \quad (17)$$

In order to solve the scattering coefficient $b_n$, the state space vector $\mathbf{D}_{1n}$ and $\mathbf{D}_{Nn}$ of the outer and inner most layers, respectively, Eq. (17) should be complemented with the continuity conditions at the fluid-PM interface

$$u_{Nnr}(r_N) = \dfrac{1}{\rho_0\omega^2}\left(a_n J'_n(k_0 r_N) + b_n H'^{(1)}_n(k_0 r_N)\right) \quad (18)$$

$$\sigma_{Nnr}(r_N) = -\left(a_n J_n(k_0 r_N) + b_n H^{(1)}_n(k_0 r_N)\right), \quad \sigma_{Nnr\theta}(r_N) = 0 \quad (19)$$

and the constraint condition on the cloak's inner surface, for which three common cases can be pursued:

$$u_{1nr}(r_0) = 0, \ u_{1n\theta}(r_0) = 0 \quad \text{for } totally\ fixed \text{ case,} \tag{20}$$

$$u_{1nr}(r_0) = 0, \ \sigma_{1nr\theta}(r_0) = 0 \quad \text{for } radially\ fixed \text{ case,} \tag{21}$$

$$\sigma_{1nr}(r_0) = 0, \ \sigma_{1nr\theta}(r_0) = 0 \quad \text{for } free \text{ case.} \tag{22}$$

After some algebraic work, the scattering coefficient for all the three boundary constraint cases can be expressed as a compact form

$$b_n = -a_n \frac{J'_n(k_0 r_N) - \rho_0 \omega^2 J_n(k_0 r_N)\chi}{H'^{(1)}_n(k_0 r_N) - \rho_0 \omega^2 H^{(1)}_n(k_0 r_N)\chi}, \tag{23}$$

where the factor $\chi$ is

$$\chi = \frac{T_{n13}T_{n44} - T_{n14}T_{n43}}{T_{n34}T_{n43} - T_{n33}T_{n44}} \quad \text{for } totally\ fixed \text{ case,} \tag{24}$$

$$\chi = \frac{T_{n12}T_{n43} - T_{n13}T_{n42}}{T_{n33}T_{n42} - T_{n32}T_{n43}} \quad \text{for } radially\ fixed \text{ case and} \tag{25}$$

$$\chi = \frac{T_{n11}T_{n42} - T_{n12}T_{n41}}{T_{n32}T_{n41} - T_{n31}T_{n42}} \quad \text{for } free \text{ case.} \tag{26}$$

In the above equations $T_{nij}$ ($i, j = 1 \sim 4$) represents elements of the transmittance matrix $\mathbf{T}_n$. With the asymptotic expansion of Hankel function for large variables, the far-field scattered pressure of Eq. (7) can be approximated in the far field as

$$p_{sc} \approx A(\theta) r^{-1/2} \exp(ik_0 r), \quad A(\theta) = \sum_{n=0}^{\infty} b_n \sqrt{\frac{2}{\pi k_0}} \exp\left(-i\left(\frac{n\pi}{2} + \frac{\pi}{4}\right)\right) \cos n\theta, \tag{27}$$

where $A(\theta)$ is the form function of the scattered pressure and the parameter $\theta$ is the azimuth angle. $A(0)$ and $A(\pi)$ are the forward and backward scattering coefficient, respectively. In order to assess the cloaking performance quantitatively, the total scattering cross section (TSCS), which accounts for the scattering in all directions, is adopted [19,39]. The TSCS is defined as the ratio between the energy scattered by the cloak and the energy normally incident onto the scatter, which is expressed as the scattering coefficient given by Eqs. (23) – (26) for each type of boundary

$$\sigma_{tot} = \frac{E_{sc}}{E_{in}} = \frac{1}{k_0 r_0} \sum_{n=0}^{\infty} (1 + \delta_{0n}) |b_n|^2. \tag{28}$$

It is noticed that the scattering coefficient of each order contributes to the TSCS. Only when the entire scattering coefficient vanishes, the cloak is perfectly invisible.

## 4. Results and Discussions

### 4.1. Shear resonance and its effect in TSCS spectrum

In the following, we first study an PM cloak with typical parameters derived from the above transformation formula. The cloak's inner and outer radii are $a$ and $b = 2a$, and the small parameter is set as $\delta = a/5$. The density $\rho$, modulus $K_r$ and $K_\theta$ in the cloak are obtained from Eq. (4) using the uniform density mapping, while $K_{r\theta}$ and $G_{r\theta}$ are deduced from the PM characteristic parameters, i.e. $K_{r\theta} = \nu(K_r K_\theta)^{1/2}$, $G_{r\theta} = \mu(K_r K_\theta)^{1/2}$. The imperfectness of PM is dictated as $\nu = 0.99$ and $\mu = 0.01$. Other choices of $\delta$ and the mapping function will not affect the main conclusion. The inner surface of the cloak is supposed to be totally fixed, and the influence of the different boundary will be addressed in the next section. In total, 21 order Bessel functions have been involved in the theoretical calculation.

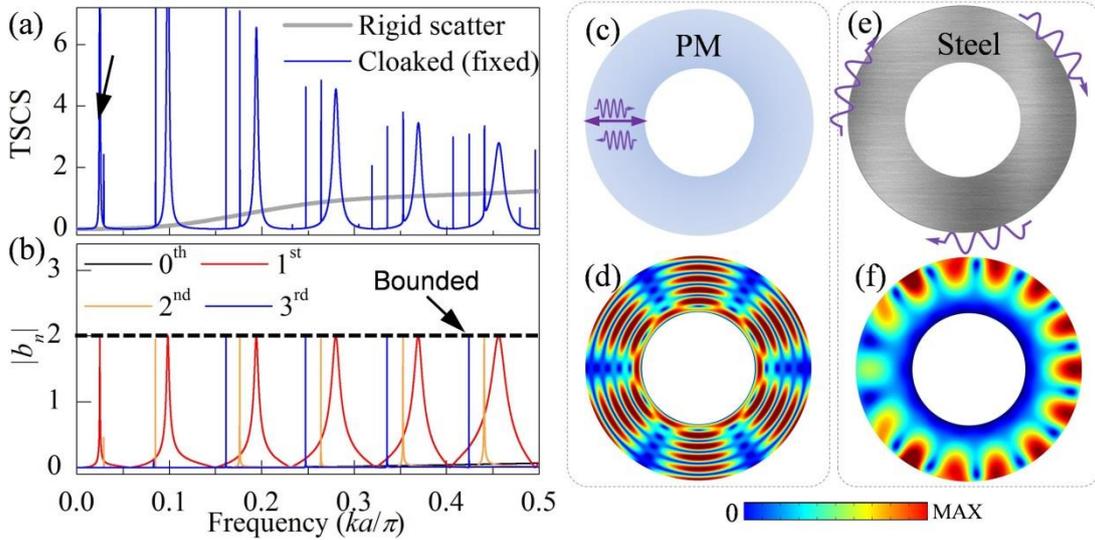

**Fig. 3.** (a) TSCS for a rigid scatter and PM cloak with totally fixed inner surface; (b) Scattering coefficient amplitude corresponding to the first four order; (c) Resonance in the unideal PM cloak is formed by shear wave along the radial direction; (d) Displacement of the PM cloak at resonance. (e) Resonance in ordinary solids is formed by whispering-gallery wave along the circumferential direction; (f) Displacement of a steel shell at resonance.

The theoretically calculated TSCS and first four order scattering amplitudes $|b_n|$ are shown in Figs. 3(a)(b). Compared to the case of bare scatter, TSCS for cloaked case has been significantly decreased for most frequencies. Sharp peak in TSCS are also observed, and they attribute to resonance shear wave due to small shear modulus [19]. We remark that the resonance mechanism here is essentially different from the conventional resonance in cylinder objects immersed in fluids, which occurs at high frequency and has been recognized for a long time [40,41]. Figure 3(e) demonstrates the mechanism for the conventional resonance, which occurs when the circumference of the cylinder is nearly an integer multiple of the wavelength. Displacement in a steel cylinder at a resonance frequency is illustrated in Fig. 3(f), and the fields are nearly periodic along the $\theta$-direction and mainly concentrated at the outer boundary. However, the resonances in the PM cloak features the shear waves along the radial direction (Fig. 3(c)), and the standing wave dominates over the entire cloak (Fig. 3(d)). This resonance happens when the shear wave travelling back and forth along the radial direction get a phase change by an integer multiple of $2\pi$. Also, different boundary condition will slightly change the resonance frequency. Based on this explanation, adjacent resonance frequency of the same order differs by $\Delta(ka/\pi) = c_T/c_0$, in which $c_T$ is the speed of shear wave in the PM. The resonance spacing can be estimated as $\Delta(ka/\pi) \approx 0.0867$ and agrees quite well with the numerical results. Another difference of the shear resonance compared to the whispering-gallery waves is that it can occur at an extremely low frequency. For instance, at the first resonance frequency (Fig. 3(a)), the wavelength in the background fluid is much larger than the cloak dimension $\lambda \approx 81a$. The low frequency resonance is dictated by the small shear modulus of the PM materials.

Another interesting thing about the resonance is that, despite the resonance peaks of the scattering coefficients are very sharp, they are always bounded $|b_n| < |a_n| \leq (2 - \delta_{0n})$. From a physical point of view, wave modes of different orders are not coupled due to a circular geometry of the cloak, therefore each scattering amplitude $|b_n|$ is always bounded by the incident amplitude $|a_n|$ of that order. Otherwise if the cloak has irregular inner or outer boundary, the scattering amplitude can be much higher than the incident wave. The bounded scattering can also be theoretically explained as follows. At the outer surface $r = r_N$ of the PM cloak, we define the $n^{th}$ order effective surface acoustic impedance

$$Z_n(k_0) = \frac{p_{nr}(r_N)}{v_{nr}(r_N)}, \tag{29}$$

where $v_{nr}(r_N)$ and $p_{nr}(r_N)$ denote the $n^{th}$ order velocity and pressure of the background fluid at $r = r_N$. With the continuity condition at the outer surface of the cloak, the effective surface impedance $Z_n(k_0)$ can be determined from the transmittance matrix Eq.(17). For instance, $Z_n$ for radially fixed inner boundary can be written as

$$Z_n(k_0) = \frac{i}{\omega}\frac{1}{\chi_1} = -\frac{i}{\omega}\frac{T_{n32}T_{n43} - T_{n33}T_{n42}}{T_{n12}T_{n43} - T_{n13}T_{n42}}. \tag{30}$$

The transmittance matrix is always real valued and therefore the effective impedance is imaginary valued. Substitute the total pressure and velocity of the background fluid into Eq.(29), we can obtain the scattering coefficient

$$b_n = -a_n \frac{k_0 J_n(k_0 r_N) - \xi(k_0) J'_n(k_0 r_N)}{k_0 H_n^{(1)}(k_0 r_N) - \xi(k_0) H_n^{'(1)}(k_0 r_N)}, \quad \xi_n(k_0) = \frac{Z_n(k_0)}{i\rho_0 c_0}. \tag{31}$$

Here, the impedance ratio $\xi_n(k_0)$ is real valued. Above solution can recover the classical result of a rigid or soft scatter. For a rigid one, the surface impedance is infinite large $\xi(k_0) = \infty$ since the surface velocity is zero, while a soft scatter implies zero impedance $\xi(k_0) = 0$ since the surface pressure is enforced to be zero. Noticing the identity $J_n(kr) Y'_n(kr) - J'_n(kr) Y_n(kr) = 2/(\pi r)$, where $Y_n(x)$ represents Bessel function of the second kind, the denominator in Eq. (31) therefore cannot be exactly zero. Thus we can conclude that the scattering coefficients cannot be infinite and indeed are always bounded by $|a_n|$,

$$|b_n| = |-a_n| \left| \frac{k_0 J_n(k_0 r_N) - \xi(k_0) J'_n(k_0 r_N)}{(k_0 J_n(k_0 r_N) - \xi(k_0) J'_n(k_0 r_N)) + i(k_0 Y_n(k_0 r_N) - \xi(k_0) Y'_n(k_0 r_N))} \right| \leq |a_n| \tag{32}$$

Eq. (32) is not only valid for PM cloaks. Indeed, the derivation implies the acoustic scattering amplitude $|b_n|$ of any cylindrical solid cannot exceed the incident amplitude $|a_n|$ of the corresponding order. According to Eq. (31), the $n^{th}$ order resonance scattering occurs when the imaginary part of the denominator is nearly zero or very small,

$$k_0 Y_n(k_0 r_N) - \xi(k_0) Y'_n(k_0 r_N) \approx 0. \tag{33}$$

Similarly, minimal scattering amplitude $b_n \approx 0$ for the $n^{th}$ order is obtained when the numerator in Eq. (31) is nearly zero $k_0 J_n(kr) - \xi(k_0) J'_n(kr) \approx 0$. The extremely small scattering can be understood as an anti-resonance phenomenon, and may be exploited to enhance the cloaking effect for a specific incident order.

*4.2. Influence of material damping and inner boundary*

In this subsection, we address the impact of material damping and inner boundary constrains on the cloaking performance. To suppress the resonance, it is natural to introduce material damping, which in practice can be realized by filling absorbing materials in the PM structures or by using lossy base materials. For illustration, the PM material is supposed to be lossy, in particular, the moduli of cloak material are multiplied by $1 + 0.005i$. The TSCS curve of damped cloak is shown in Fig. 4(a). It is found that most narrow peaks are significantly suppressed except the wide ones corresponding to the first order resonances. Figs. 4(b) and 4(c) further show the results for the radially fixed and free boundary on the inner surface. Among the three boundary conditions, the radially fixed one provides the most preferable concealing (Fig. 4(b)); all the resonant peaks are narrow and can be well suppressed by damping. For the free case (Fig. 4(c)), a very wide resonance extends through the entire frequency range and significantly damages the cloaking effect. This resonance is induced by zeroth order wave and is independent of the shear modulus. Such a resonance has also observed in the anisotropic density acoustic cloak [42].

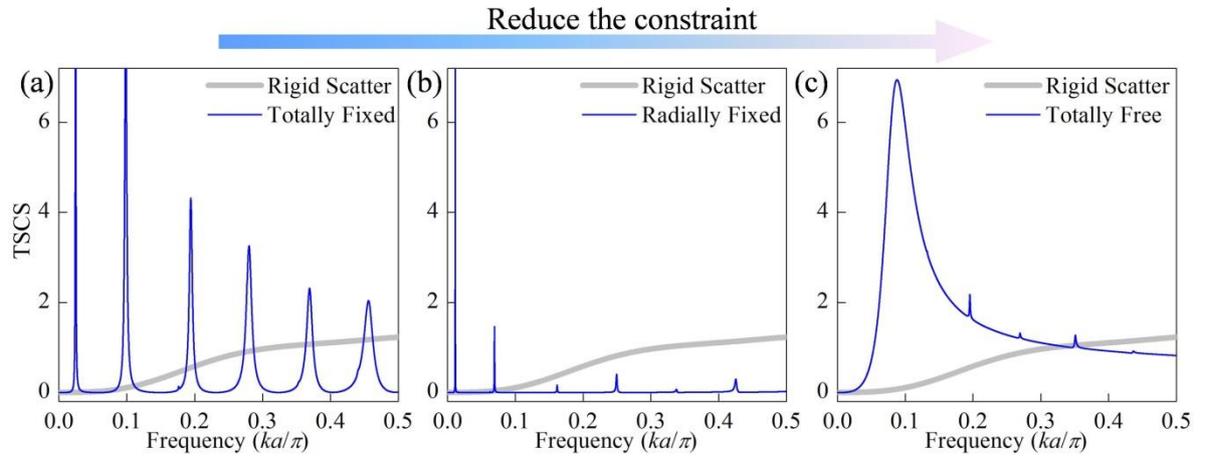

**Fig. 4.** TSCS for a damped cloak (damping ratio 0.5%) with (a) Totally fixed inner surface; (b) Radially fixed inner surface; (c) Free inner surface.

*4.3. Practical inner boundary by a liner solid shell*

In reality, the most convenient inner boundary for PM cloak is the free one. However, its cloaking effect is the worst due to the strong zeroth order scattering. To this end, we propose to support the PM cloak by a homogeneous solid shell on its inner surface (Fig. 5(a)). The cloak and shell are supposed to be perfected bonded and the inner surface of the shell is left free. In doing so, the constraint strength of the inner side of the PM can be tuned continuously by varying the supporting shell thickness. This not only provides a very practical inner boundary but also offers further insights for the boundary effect.

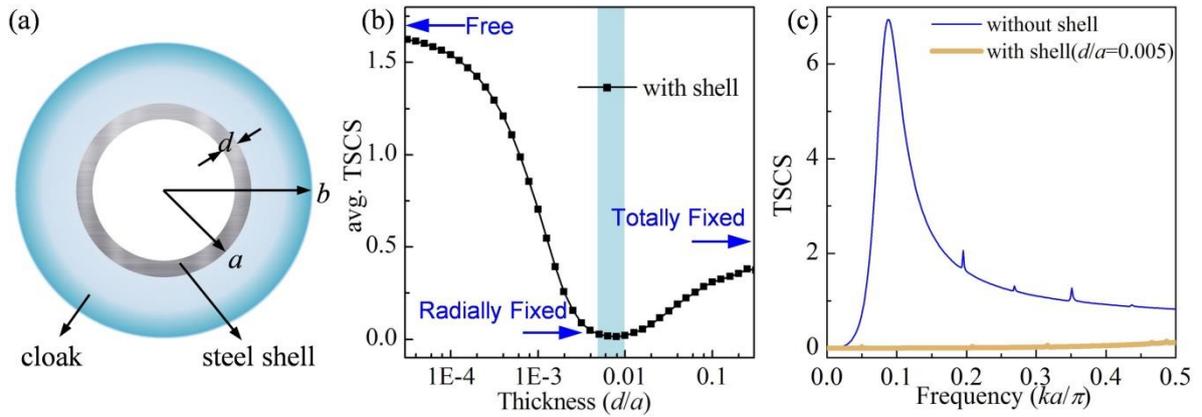

**Fig. 5.** Cloak supported by a traction free inner shell; (a) Schematic configuration; (b) Averaged TSCS with respect to the shell thickness; (c) TSCS for cloak without/with shell.

The cloak parameters and material damping are the same as the previous ones, and the shell with thickness $d$ is chosen to be of steel (density $\rho_{steel}$ = 7800 kg/m$^3$, Young's modulus $E_{steel}$ = 220GPa and Poisson's ratio $v_{steel}$ = 0.28). Figure 5(b) plots the average TSCS as a function of the shell thickness. The TSCS is estimated in an average sense over frequency range $ka/\pi = 0 - 0.5$. For the cloak supported by very thinner shell, the averaged TSCS is close to that of a cloak with free inner boundary. By increasing the shell thickness, an excellent cloaking performance is observed for the range of $d/a = 0.005 - 0.01$, as if the inner boundary is radially fixed. Further increasing the shell thickness will increase the constraint strength and the TSCS tends to be the result of totally fixed case. The TSCS for the cloak without/with shell is also plotted in Fig. 5(c). For the cloak supported with an appropriate shell, all the resonances are well suppressed by the material damping as one would expect.

To further verify the above strategy of virtually tuning the inner surface boundary of the PM acoustic cloak. We improve our previously designed microstructure cloak with a supporting steel shell attached to its inner surface (Fig. 6(a)). The PM cloak is completely constructed from PM unit cells (Fig. 6(b)) by optimizing geometry parameters to match the required PM properties. The density and bulk modulus (Fig. 6(c)) is a good approximation to the required continuous material parameters. One can refer to our paper [19] for the detail microstructure design procedure. The shell thickness is the same as the above $d/a = 0.005$ and the matrix material has a damping factor 0.5%. Figure 6(d) plots the simulated TSCS for the microstructure cloak with/without shell. A very good broadband cloaking effect is obtained for the cloak with a supporting shell, and the pressure fields in Figs. 6(e)(f) clearly demonstrate the prominent stealth effect.

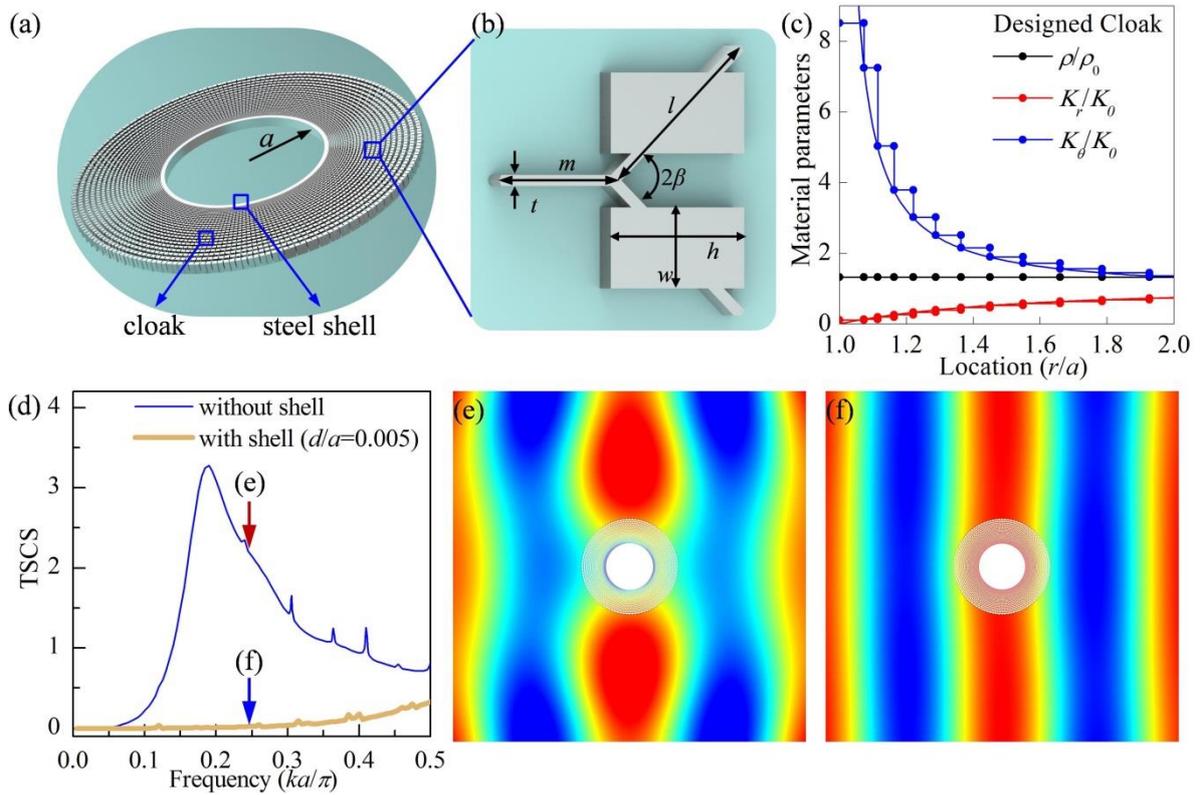

**Fig. 6.** Verification of the proposed boundary scheme through microstructure PM cloak; (a) Microstructure PM cloak with an inner shell; (b) PM unit cell; (c) Material parameter distribution; (d) TSCS for cloak without/with shell; (e)(f) Pressures for cloak without/with shell at $ka/\pi = 0.25$.

## 5. Conclusions

In this paper, a theoretical model is developed to study the scattering of cylindrical acoustic cloak composed of imperfect PM with shear rigidity, and the impact of material parameters, damping and inner boundary are systematically investigated. It is found that, shear rigidity of the PM introduces intense resonances at low frequency range and disturbs the broadband effectiveness. The resonance is explained as shear wave resonance in the radial direction, and is different to the traditional whispering-gallery resonance along the circumference. Three boundary conditions on the cloak inner surface, i.e. radially fixed, totally fixed and free cases, are examined. In general, the most preferred broadband cloaking is achieved with radially fixed boundary. For the totally fixed boundary, the first order resonance shows quite large width, while obvious zeroth order scattering is observed over the entire low frequency range for the free boundary. To overcome the difficulty in realizing a radially constrained inner boundary, we further propose to attach a solid shell to the inner surface of the cloak. It is interesting to find that, the constraint strength on the cloak inner surface can be continuously tuned to range over the three boundary types by varying the shell thickness. This boundary scheme is further verified by simulation of a PM cloak with microstructure implementation. This work is instructive for the PM cloak implementation taking into account the PM imperfectness, inner constraints and broadband cloaking effectiveness.

**Acknowledgements**

We are grateful to Andrew Norris for discussing on resonance. This work was supported by the National Natural Science Foundation of China (grant numbers 11372035, 11472044, 11632003, 11802017), Postdoctoral Innovation Talents Support Program (No. BX20180040), and the 111 Project (No. B16003).

**Declarations of interest**: none

## References

[1] Pendry JB, Schurig D, Smith DR. Controlling electromagnetic fields. Science 2006;312(5781):1780-2.

[2] Leonhardt U. Optical conformal mapping. Science 2006;312(5781):1777-80.

[3] Chen HY, Chan CT. Acoustic cloaking in three dimensions using acoustic metamaterials. Appl Phys Lett 2007;91(18):183518.

[4] Cummer SA, Schurig D. One path to acoustic cloaking. New J Phys 2007;9(3):45.

[5] Norris AN. Acoustic cloaking theory. Proceedings of the Royal Society A: Mathematical, Physical and Engineering Sciences 2008;464(2097):2411-34.

[6] Cheng Y, Yang F, Xu JY, Liu XJ. A multilayer structured acoustic cloak with homogeneous isotropic materials. Appl Phys Lett 2008;92(15):151913.

[7] Christensen J, de Abajo FJG. Anisotropic metamaterials for full control of acoustic waves. Phys Rev Lett 2012;108(12):124301.

[8] Liu AP, Zhu R, Liu XN, Hu GK, Huang GL. Multi-displacement microstructure continuum modeling of anisotropic elastic metamaterials. Wave Motion 2012;49(3):411-26.

[9] Milton GW, Willis JR. On modifications of Newton's second law and linear continuum elastodynamics. Proceedings of the Royal Society A: Mathematical, Physical and Engineering Sciences 2007;463(2079):855-80.

[10] Torrent D, Sanchez-Dehesa J. Anisotropic Mass Density by Radially Periodic Fluid Structures. Phys Rev Lett 2010;105(17):174301.

[11] Popa BI, Wang W, Konneker A, Cummer SA, Rohde CA. Anisotropic acoustic metafluid for underwater operation. J Acoust Soc Am 2016;139(6):3325-31.

[12] Popa B, Zigoneanu L, Cummer SA. Experimental Acoustic Ground Cloak in Air. Phys Rev Lett 2011;106(25):253901.

[13] Zigoneanu L, Popa B, Cummer SA. Three-dimensional broadband omnidirectional acoustic ground cloak. Nat Mater 2014;13(4):352-5.

[14] Milton GW, Cherkaev AV. Which elasticity tensors are realizable? Journal of Engineering Materials and Technology 1995;117(4):483-93.

[15] Gokhale NH, Cipolla JL, Norris AN. Special transformations for pentamode acoustic cloaking. J Acoust Soc Am 2012;132(4):2932-41.

[16] Chen Y, Liu X, Hu G. Design of arbitrary shaped pentamode acoustic cloak based on quasi-symmetric mapping gradient algorithm. J Acoust Soc Am 2016;140(5):L405-9.

[17] Zhao A, Zhao Z, Zhang X, Cai X, Wang L, Wu T, et al. Design and experimental verification of a water-like pentamode material. Appl Phys Lett 2017;110(1):11907.

[18] Cai X, Wang L, Zhao Z, Zhao A, Zhang X, Wu T, et al. The mechanical and acoustic properties of two-dimensional pentamode metamaterials with different structural parameters. Appl Phys Lett 2016;109(13):131904.

[19] Chen Y, Liu XN, Hu GK. Latticed pentamode acoustic cloak. Sci Rep-Uk 2015;515745.

[20] Tian Y, Wei Q, Cheng Y, Xu Z, Liu XJ. Broadband manipulation of acoustic wavefronts by pentamode metasurface. Appl Phys Lett 2015;107(22):221906.

[21] Layman CN, Naify CJ, Martin TP, Calvo DC, Orris GJ. Highly Anisotropic Elements for Acoustic Pentamode Applications. Phys Rev Lett 2013;111(2):24302.

[22] Cipolla J, Gokhale N, Norris A, Nagy A. Design of inhomogeneous pentamode metamaterials for minimization of scattering. J Acoust Soc Am 2011;130(4):2332.

[23] Scandrett CL, Boisvert JE, Howarth TR. Broadband optimization of a pentamode-layered spherical acoustic waveguide. Wave Motion 2011;48(6):505-14.

[24] Scandrett CL, Boisvert JE, Howarth TR. Acoustic cloaking using layered pentamode materials. J Acoust Soc Am 2010;127(5):2856-64.

[25] Hladky-Hennion AC, Vasseur JO, Haw G, Croenne C, Haumesser L, Norris AN. Negative refraction of acoustic waves using a foam-like metallic structure. Appl Phys Lett 2013;102(14):144103.

[26] Chen Y, Zheng M, Liu X, Bi Y, Sun Z, Xiang P, et al. Broadband solid cloak for underwater acoustics. Phys Rev B 2017;95(18010418).

[27] Zhaoyong S, Han J, Yi C, Zhen W, Jun Y. Design of an underwater acoustic bend by pentamode metafluid. J Acoust Soc Am 2018;3(142):1029-34.

[28] Martin A, Kadic M, Schittny R, Bückmann T, Wegener M. Phonon band structures of three-dimensional pentamode metamaterials. Phys Rev B 2012;86(15):155116.

[29] Kadic M, Bückmann T, Stenger N, Thiel M, Wegener M. On the practicability of pentamode mechanical metamaterials. Appl Phys Lett 2012;100(19):191901.

[30] Kadic M, Bückmann T, Schittny R, Wegener M. On anisotropic versions of three-dimensional pentamode metamaterials. New J Phys 2013;15(023029).

[31] Kadic M, Bückmann T, Schittny R, Gumbsch P, Wegener M. Pentamode metamaterials with independently tailored bulk modulus and mass density. Phys Rev Appl 2014;2(5):54007.

[32] Huang Y, Lu XG, Liang GY, Xu Z. Pentamodal property and acoustic band gaps of pentamode metamaterials with different cross-section shapes. Phys Lett a 2016;380(13):1334-8.

[33] Amendola A, Benzoni G, Fraternali F. Non-linear elastic response of layered structures, alternating pentamode lattices and confinement plates. Composites Part B 2017;115117-23.

[34] Fraternali F, Amendola A. Mechanical modeling of innovative metamaterials alternating pentamode lattices and confinement plates. J Mech Phys Solids 2017;99259-71.

[35] Schittny R, Bückmann T, Kadic M, Wegener M. Elastic measurements on macroscopic three-dimensional pentamode metamaterials. Appl Phys Lett 2013;103(23):231905.

[36] Smith JD, Verrier PE. The effect of shear on acoustic cloaking. P Roy Soc a-Math Phy 2011;467(2132):2291-309.

[37] Chen WQ, Bian ZG, Ding HJ. Three-dimensional vibration analysis of fluid-filled orthotropic FGM cylindrical shells. Int J Mech Sci 2004;46(1):159-71.

[38] Hasheminejad SM, Rajabi M. Acoustic resonance scattering from a submerged functionally graded cylindrical shell. J Sound Vib 2007;302(1-2):208-28.

[39] Titovich AS, Norris AN. Tunable cylindrical shell as an element in acoustic metamaterial. J Acoust Soc Am 2014;136(4):1601-9.

[40] Norris AN. Resonant acoustic scattering from solid targets. The Journal of the Acoustical Society of America 1990;88(1):505-14.

[41] Flax L, Dragonette LR, Überall H. Theory of elastic resonance excitation by sound scattering. J Acoust Soc Am 1978;63(3):723-31.

[42] Cheng Y, Liu XJ. Resonance effects in broadband acoustic cloak with multilayered homogeneous isotropic materials. Appl Phys Lett 2008;93(7):71903.